\documentclass[pra,aps,twocolumn,eqsecnum,showpacs]{revtex4}
\usepackage{mathtext}
\usepackage[T2A]{fontenc}
\usepackage[cp1251]{inputenc}
\usepackage{epsf}
\usepackage{psfrag}
\usepackage{graphicx}
\usepackage{amssymb,amsmath}

\begin{document}

\title{Dispersion of the dielectric permittivity of dense and cold atomic gases}
\author{Ya. A. Fofanov${}^{1}$, A. S. Kuraptsev${}^{2}$,   and I. M. Sokolov${}^{1,2}$\\
{\small $^{1}$Institute for Analytical Instrumentation, Russian Academy of
Sciences, 190103, St.-Petersburg, Russia }\\
{\small $^{2}$Department of Theoretical Physics, State Polytechnic
University, 195251, St.-Petersburg, Russia }\\
M. D. Havey\\
{\small Department of Physics, Old Dominion University, Norfolk, VA 23529}}

\sloppy



\begin{abstract}
On the basis of general theoretical results developed previously in JETP 112, 246 (2011) we analyze the atomic polarization created by weak monochromatic light in an optically thick, dense and cold atomic ensemble. We show that the amplitude of the polarization averaged over a uniform random atomic distribution decreases exponentially beyond the boundary regions. The phase of this polarization increases linearly with increasing penetration into the medium. On these grounds, we determine numerically the wavelength of the light in the dense atomic medium, its extinction coefficient, and the complex refractive index and dielectric constant of the medium. The dispersion of the permittivity is investigated for different atomic densities. It is shown that for dense clouds, the real part of the permittivity is negative in some spectral domains.
\end{abstract}


\maketitle

\section{Introduction}
Improvements in techniques for cooling of atomic gases in atomic traps make their use very promising for practical applications in various areas of fundamental science and technology such as metrology, development of frequency standards, and quantum information problems \cite{3,4,5,6,7,8,9,10,11}. The largest number of applications envisioned for cold and ultracold atomic
ensembles have at their foundation the interaction between the medium and electromagnetic radiation. This interaction also underlies many methods for diagnostics of the states of these ensembles. Recently, dense atomic clouds, in which the average interatomic distances are comparable with the optical wavelength, have attracted much attention. This interest is in part generated by such fundamental and practically important physical effects as Anderson (strong) localization of light \cite{24,25,26,27} and lasing in a disordered medium \cite{28,29,30,31,32} which can take place in dense ensembles. The aim of recent studies in this field was to observe these effects experimentally and to describe them theoretically.

One challenging problem in the area of disordered atomic systems is that the studied atomic ensembles normally consist of a large number of atoms in samples that are produced with a low duty cycle.  The larger number of atoms is required in order to obtain sufficient signal to noise to study the subtle effects of interest.  Such experiments require realistic modeling in order to extract the essential physics of the observed processes.  However, it is challenging to treat these problems as a multi-atom scattering process, and such studies have been limited to several thousand atoms
\cite{SKKH09,SKH11,BWHSK1}; this should be compared with the characteristic $10^{6}$ atom-sized samples of recent experiments. Alternative theoretical approaches, even if approximate in nature, are then desirable.

The present paper is devoted to the theoretical description of optical properties of dense and cold atomic clouds.  The problem of a dense atomic ensemble belongs obviously to the field of macroscopic electrodynamics. The main approach here is based on usage of such averaged characteristics as the field strength and atomic polarization. The key point in a macroscopic approach is in finding the susceptibility or dielectric constant of the dense ensemble. The influence of density of the medium on its susceptibility can be analyzed on the basis of the idea of a local field and, following from it, the Lorentz-Lorenz formula \cite{BW}.   This formula is sufficient to solve completely the problem of the dependence on density only if the difference between the polarizability of a free atom and its polarizability in the medium can be neglected \cite{Kaiser1}. As we will show below, for the considered cold atomic ensemble this is not the case. The resonant dipole-dipole interatomic interaction causes atomic level shifts and broadening and thereby essentially modifies the atomic polarizability. An explicit analytical expression for the susceptibility, which takes into account this modification, was obtained earlier in \cite{SKKH09}. The calculation in \cite{SKKH09} was based on the relevant macroscopic statistical description of the polarization response of the medium to an external field. Part of the approximations made in \cite{SKKH09} are valid only for relatively low density ensembles and thus the corresponding results have a restricted range of applicability. Constitutive relations connecting atomic polarization and an external field can be obtained consistently only in the framework of a microscopic approach based on the notion of the discrete structure of matter consisting of separate atoms.

A consistent microscopic approach has been already applied for analysis of influence of interatomic interactions on spontaneous decay of an impurity atom embedded in a dielectric \cite{Berman,Berman1,Berman2,Berman3}. Quite a number of works were devoted to collective decay in dense homogeneous multiatomic media and to properties of spontaneous emission of such media initially excited by a weak external field (see, for example \cite{Lehmberg70,Fl,RO,Pin1,Akk,Akk1,MK07,FrM,FrM1,Sw1,Scully,Sw} and references therein). In these works the main attention was focused on the influence of the density of the ensemble on its afterglow, i.e. on secondary radiation. In the present paper we consider the influence of interatomic interaction on the properties of the ensemble itself. We study the spatial distribution of atomic polarization created by weak monochromatic light in a cold atomic ensemble. We show that in the case of a uniform random atomic distribution amplitude of polarization averaged over space configurations decreases exponentially beyond the boundary regions. Its phase increases linearly with distance into the medium. On this ground we determine numerically the wavelength of the light in the dense atomic clouds, its extinction coefficient as well as complex index of refraction and dielectric constant of the medium. We also analyze the dispersion of the permittivity for different atomic densities. Note that similar problems arise in classical electrodynamics when studying light scattering from a medium with random dielectric inclusions \cite{Pellegrin1}. Such inclusions strongly influence light propagation in such media, giving rise manifold internal scattering and essentially modifying its averaged dielectric constant.

An important feature of the present work is in taking into account the polarization properties of light. Nearly all the mentioned above papers on multiatomic systems used only a two-level model for the atoms. This prevents a correct consideration of the light polarization, adequate description of the resonance dipole-dipole interaction at small distances and, as a consequence, correct calculation of shifts and broadenings of atomic levels. We also do not use a model of an averaged continuous medium in our calculations. That is, the influence on the intrinsic spatial disorder of the atoms in the ensemble is considered.  As a specific illustration of this, recent approaches to atomic physics-based localization studies have considered systems of reduced dimensionality.  One way to achieve this for light localization is to optically create a quasi one dimensional system through modification of spatially larger samples.  Such optical channels, with wavelength-scale transverse dimensions, can be created through quantum optical techniques based on electromagnetically induced transparency, for instance.  Alternatively, a strongly focused far-off-resonance laser can generate a type of optical wave guide through the dense sample, allowing quasi one dimensional localization for a much weaker, but near-resonance probe beam.  Theoretical modeling the average properties of such generated optical wave guides, essential to interpretation of experiments, may be done using the effective optical responses of the resulting medium, as we discuss in the current paper.

The remainder of this paper is organized as follows. In Section 2 we describe our basic physical assumptions and the calculational approach. Section 3 presents results of numerical simulations.  We conclude with a brief synopsis of the results, highlighting the main points of the present report.

\section{Basic assumptions and approaches}
Consider the temporal dynamics of a system consisting of $N+1$ motionless atoms. Let $N$ atoms form the cloud. These atoms are identical and have a ground state $J=0$ separated by the frequency $\omega_a$ from an excited $J=1$ state. The natural linewidth of this state is $\gamma$. One atom is located far from the cloud and has the same $J=0 \leftrightarrow J=1$  structure of levels but a different transition frequency $\omega_s$ and a different decay constant $\gamma_s$. We will assume that initially all atoms of the cloud are in the ground state and the separated atom, which we will refer to as a source atom or simply the source, is in a coherent state which is a superposition of the ground and a small admixture of the excited state. In the course of spontaneous decay such an atom creates an electromagnetic field which is a superposition of vacuum and a small admixture of a one photon state.  As is known, this superposition approximates a weak coherent state of the field with good accuracy. Under the influence of the field, the atoms of the cloud are excited and in due course emit secondary radiation which can be absorbed by other atoms of the cloud. The process of manyfold photon exchange determines the dipole-dipole interatomic interaction and manifests itself in such phenomena as spontaneous decay modification, collective atomic state formation and so on.

The microscopic description of dynamics of the considered ensemble is based on the non stationary Schrodinger equation for the wave function $\psi $ of the joint system consisting of atoms and the field generated in the process of the evolution
\begin{equation} i\hbar \frac{\partial \psi
}{\partial t}=H\psi ,  \label{1}
\end{equation}
The Hamiltonian of the system $H$ can be presented as the sum of the Hamiltonians of the free atoms and the free field $H_0$, and
the operator $V$ of their interaction. In the dipole approximation used here, we have
\begin{equation}
V=-\sum_{a}\mathbf{d}^{(a)}\mathbf{E}(\mathbf{r}_{a}),  \label{3}
\end{equation}%
\begin{eqnarray}
\mathbf{E}(\mathbf{r})&=&\mathbf{E}^{(+)}(\mathbf{r})+\mathbf{E}^{(-)}(\mathbf{%
r})=\label{4}\\
&=&i\sum\limits_{\mathbf{k},\alpha }\sqrt{\frac{2\pi \hbar \omega _{k}}{\cal V}}%
\mathbf{e}_{\mathbf{k}\alpha }a_{\mathbf{k}\alpha }\exp (i\mathbf{kr})+h.c.
\nonumber
\end{eqnarray}%
where $\mathbf{E}^{(\pm )}$ are the operators of the positive and negative frequency components of the field;
$a_{\mathbf{k}\alpha }$ is the photon annihilation operator in a mode with wave vector $\mathbf{k}$ and polarization $\alpha ;$ $\cal V$ is the quantization volume; $\mathbf{d}^{(a)}$  is the dipole moment operator of the atom $a$, $\mathbf{e}_{\mathbf{k}\alpha }$ are polarization unit vectors.

We will seek the wave function  $\psi $ as an expansion in a set of eigenstates $\left\{ |l\rangle \right\} $ of the
operator $H_{0}$:
\begin{equation} \psi =\sum_{l}b_{l}(t)|l\rangle .
\label{6}
\end{equation}
Here, the subscript $l$ defines the state of all atoms and the field.

The key simplification of the approach employed is in restriction of total number of states $|l\rangle$ taken into account. We will calculate all radiative correction up to the second order of the fine structure constant. In this case we can consider only the following states (see \cite{Stephan64})
\begin{eqnarray}
\psi _{g} &=&|g,g,...g\rangle \otimes |\mathbf{k}\alpha \rangle
 \label{8.1} \\
\psi _{g'} &=&|g,g,...g\rangle \otimes |vac \rangle
\label{8.2} \\
\psi _{e_{a}} &=&|g,g,...g,e,g,...g\rangle \otimes |vac\rangle
 \label{8.3} \\
\psi _{e_{a}e_{b}} &=&|g,...g,e,g...g,e,g,...g\rangle \otimes |\mathbf{k}%
\alpha \rangle  \label{8.4}
\end{eqnarray}
In the rotating wave approximation it is enough to take into account only the states (\ref{8.1}) and (\ref{8.3}). States without excitation both in atomic and field subsystem (\ref{8.2}) allow us to describe coherent states of the source atom. Non resonant states with two excited atoms and one photon (\ref{8.4}) are necessary for a correct description of the dipole-dipole interaction at short interatomic distances. Note that, in considered case, there are three excited states for each atom  $e=|J,m\rangle$, which differ by the value of angular momentum projection $m=-1,0,1$. Therefore, the total number of one-fold excited states (\ref{8.3}) is $3(N+1)$.

Equation (\ref{1}) should be supplemented by an initial condition. According to our previous discussion, we will consider the case when initially the field is in a vacuum state, all atoms of the cloud are in the ground state and the source atom which we denote by index $s$  is in superposition of ground and one of the excited states $|J,m\rangle$. Designating  the corresponding amplitudes as  $b'_0$ and $b_0$, we can write
\begin{equation} \psi(0) =b'_0|g'\rangle +b_0|e_{s0}\rangle ,
\label{9}
\end{equation}
where the index $e_{s0}$ corresponds to the one of the three possible states of atom $s$ which is populated in the initial moment of time.

In the framework of the assumptions made, the amplitude of state $\psi _{g^{\prime }}=|g'\rangle$ does not change during the evolution of the system $b_{g^{\prime }}(t)=b'_{0}$, because transitions to this state from other states taken into account are impossible. The transition from (\ref{8.2}) to any of the states  is also impossible.

To determine all other amplitudes we have to solve the set of equations which follows from  (\ref{1}). In spite of the performed restriction of the number of states, this set of equations is infinite because of the infinite number of degrees of freedom of the field. We can however exclude amplitudes of states with one photon and obtain a finite closed system of equations for $b_{e}(t)\equiv b_{e_{a}}(t);\, a \neq s $. For Fourier components $b_{e}(\omega)$ we have (at greater length see \cite{SKH11})
\begin{equation}
\sum\limits_{e^{\prime }\neq s}\left[ (\omega -\omega _{a})\delta
_{ee^{\prime }}-\Sigma _{ee^{\prime }}(\omega )\right] b_{e^{\prime
}}(\omega ) =\Lambda_{es}(\omega ).
\label{19}
\end{equation}

Matrix elements $\Sigma _{ee^{\prime }}(\omega )$ for $e$ and $e^\prime$ corresponding to different atoms describe excitation exchange between these atoms. Assuming that in state $\psi _{e^{\prime }}$  and $\psi _e$ atoms $b$  and $a$ are excited correspondingly, in the framework of the pole approximation (see \cite{MK74}), we have
\begin{eqnarray}
&&\Sigma _{ee^{\prime }}(\omega )=\sum\limits_{\mu ,\nu}
\frac{\mathbf{d}_{e_{a};g_{a}}^{\mu }\mathbf{d}_{g_{b};e_{b}}^{\nu }}{\hbar r^{3}}\times \\&& \left[ \delta _{\mu \nu }\left(
1-i\frac{\omega _{a}r}{c}-\left( \frac{\omega _{a}r}{c}\right) ^{2}\right)
\exp \left( i\frac{\omega _{a}r}{c}\right) \right. -
\notag  \\&&
\left. -\dfrac{\mathbf{r}_{\mu }\mathbf{r}_{\nu }}{r^{2}}\left( 3-3i\frac{\omega _{a}r}{c}-\left( \frac{\omega _{a}r}{c}\right) ^{2}\right) \exp
\left( i\frac{\omega _{a}r}{c}\right) \right] .
\notag
\end{eqnarray}
Here $\mathbf{r}_\mu$ is the projection of the vector $\mathbf{r}=\mathbf{r}_{a}-\mathbf{r}_{b}$ on the axis of the chosen coordinate system and  $r=|\mathbf{r}|$ is the separation between atoms $a$ and $b$.

If $e$ and $e^\prime$ correspond to excited states of one atom then $\Sigma _{ee^{\prime }}(\omega )$ differs from zero only for $e=e^\prime$ ($m=m^\prime$). In this case $\Sigma _{ee}(\omega )$ determines the Lamb shift and the decay constant of corresponding excited state. Including Lamb shifts in the transition frequency $\omega _{a}$ we get
\begin{equation} \Sigma _{ee}(\omega )=-i\gamma _{a}/2.  \label{27}
\end{equation}

The term $\Lambda_{es}(\omega )$ in the right-hand side of Eq. (\ref{19}) describes excitation of the cloud atoms by the radiation of the source.
Assuming that the size of the atomic ensemble is negligible compare with the distance from it to the source, and neglecting the secondary excitation of the source atom $s$ by reradiation from the cloud, we have
\begin{eqnarray}
\Lambda_{es}(\omega )&=&\frac{ib_{0}}{(\omega -\omega _{s}+i\gamma_s /2)}\widetilde{\Sigma}_{es}(\omega ),  \label{27.1}\\
\widetilde{\Sigma} _{es }(\omega )&=&-\sum\limits_{\mu ,\nu}
\frac{k^{2}\mathbf{d}_{e;g}^{\mu }\mathbf{d}_{g_{s};e_{s}}^{\nu
}}{\hbar r_{s}}\left[ \delta _{\mu \nu }-\frac{\mathbf{k}_{\mu
}\mathbf{k}_{\nu }}{k^{2}}\right]\times\nonumber\\ &&\times \exp \left( ikr_{s}+i\mathbf{kr}_{e}\right).
\label{27.2}
\end{eqnarray}
Here $\mathbf{k=}\omega \mathbf{n/}c$. Relation (\ref{27.2}) is written in a coordinate frame originating at some point inside the cloud; $\mathbf{r}_{e}$ are radii locating the atoms; $\mathbf{n}$ is a unit vector oriented from the source to the cloud.   In obtaining the expression for $\widetilde{\Sigma} _{es }(\omega )$ we took into account the non-applicability of the pole approximation because of the large separation between the cloud and source, we used the rotating wave approximation for the same reason and kept only one term which decreases most slowly with $r_{e}$.  All these factors generate the differences between $\widetilde{\Sigma}$ and the elements of matrix $\Sigma$.

Knowledge of explicit expressions for $\Lambda_{es}(\omega )$ and $\Sigma _{ee^{\prime }}(\omega )$ allows us to determine the amplitudes of all one-fold excited states (\ref{8.3}). Note that system (\ref{19}) can be reduced to an integral equation by using the continuous medium approximation. This significantly simplifies the solution of the problem for a two-level atom system \cite{FrM,FrM1,Sw1,Scully,Sw}. Moreover, in this case, even an analytic solution is possible for spatially homogeneous spherical clouds. This solution neglects, however, the important properties of real physical systems, and therefore we will solve the linear system (\ref{19}) numerically. In a numerical solution we can correctly describe all polarization effects and take into account the random inhomogeneities of the medium.

Introducing the inverse matrix which, as shown in \cite{SKKH09}, is a resolvent operator of the considered multi atomic cloud
\begin{eqnarray}
R_{ee^{\prime }}(\omega )=\left[ (\omega -\omega _{a})\delta _{ee^{\prime }}-\Sigma
_{ee^{\prime }}(\omega )\right] ^{-1},  \label{21.1}
\end{eqnarray}
we can write the solution of the system (\ref{19}) as follows
\begin{eqnarray}
b_{e}(\omega ) =\sum\limits_{e^{\prime }\neq
s}R_{ee^{\prime }}(\omega )\Lambda _{e^{\prime }s}(\omega ).  \label{21}
\end{eqnarray}
For amplitude $b_{e}(t)$ we get
\begin{eqnarray}
b_{e}(t) &=&\int\limits_{-\infty }^{\infty }\dfrac{id\omega }{2\pi }\frac{%
b_{0}\exp (-i\omega t)\sum\limits_{e^{\prime }\neq s}R_{ee^{\prime }}(\omega
)\tilde{\Sigma} _{e^{\prime }s}(\omega )}{\omega -\omega _{s}+i\gamma_s /2}.\nonumber \\ \label{22}
\end{eqnarray}%

This relation give us the possibilities to find the distribution of excited states at any instant of time. In this work we are interested in the spatial distribution under quasi static conditions. Such conditions can be realized if decay times of all collective states of the dense atomic ensemble are much less than the decay time of the source atom  $s$.

Let us consider the relation (\ref{22}) for a time interval much less that $\gamma^{-1}_s$ but larger than the mentioned collective relaxation times. Formally, relations for the quasi steady state regime can be obtained by two limiting processes. First we should pass to the limit $\gamma_s\rightarrow 0$ and then to $t\rightarrow \infty$. Realizing these limiting processes and taking into account that $\underset{\gamma_s\rightarrow 0}{lim}(\omega-\omega _{s}+i\gamma_s /2)^{-1}=\varsigma (\omega
-\omega _{s})$, where $\varsigma (x)$ is a singular function and that $\underset{t\rightarrow +\infty }{lim}\varsigma \left( \omega
-\omega _{s}\right) \exp (-i\omega t)=-2\pi i\delta (\omega-\omega _{s})\exp (-i\omega _{s}t )$ we get
\begin{eqnarray}
b_{e}(t)=b_{0}\exp (-i\omega _{s}t)\sum\limits_{e^{\prime }\neq s}R_{ee^{\prime }}(\omega _{s})\tilde{\Sigma} _{e^{\prime }s}(\omega _{s}).\nonumber \\
\end{eqnarray}%

By using $b_{e}(t)$ we can obtain amplitudes of all states taken into account in our calculations (see \cite{SKH11}) and consequently, the wave function of the considered system.  Among other things, this allows calculation of the  polarization as the averaged dipole moment of unit volume of the atomic ensemble. For a given projection $\mu$ of the polarization vector we have
\begin{equation}
\mathcal{P}_\mu(\mathbf{r},t)=\frac{1}{\Delta V}\sum_{a\in \Delta
V}\langle\hat{d}^{(a)}_{\mu}\rangle.   \label{2.7}
\end{equation}
Here $\hat{d}^{(a)}_{\mu}$ is the operator of the corresponding projection of the dipole moment of atom $a$. The summation in (\ref{2.7}) is made over all atoms located in a mesoscopic volume $\Delta V$ near the point $\mathbf{r}$. Quantum-mechanical averaging is performed over the wave function of the system.

In analyzing the polarization it is convenient to select positive and negative frequency parts and use a basis of circular polarization  ($\mu=0,\pm 1$):
\begin{equation}
\mathcal{P}_\mu(\mathbf{r},t)=\mathcal{P}_\mu^{(-)}(\mathbf{r},t)+\mathcal{P}_\mu^{(+)}(\mathbf{r},t).
\label{2.8}
\end{equation}
Using the known wave function and taking into account the short life time of the nonresonant virtual states with two excited atoms and consequently its small contribution to the polarization, we find
\begin{eqnarray}
\mathcal{P}_\mu^{(+)}(\mathbf{r},t)&=&\mathcal{P}_\mu^{(+)}(\mathbf{r})\exp (-i\omega _{s}t);\label{2.9}  \\
\mathcal{P}_\mu^{(+)}(\mathbf{r})&=&\frac{b'^*_{0}b_{0}}{\Delta V}\sum_{a\in \Delta V}\notag
\sum\limits_{e_b}R_{e^m_ae_b}(\omega _{s})\tilde{\Sigma} _{e_bs}(\omega _{s}).
\end{eqnarray}
The additional index $m$ at $e_a$ means that under summation we have to include only those states $e^m_a$ of atom $a$ which give contributions to the corresponding projection of the polarization vector. In the basis of circular polarization such contribution comes only from one Zeeman sublevel with $m=\mu$. Due to the optical isotropy of the atomic ensemble, the  orientation of the atomic polarization vector coincides with the orientation of the polarization of light exciting the atoms. The latter, in turn, depends on the specific Zeeman sublevel $m_s$ of the source atom which was excited initially. In the case when the quantization axis coincides with the vector $\mathbf{n}$ the configurationally averaged atomic polarization has only one nonzero projection $\mu=m_s$. Thus to determine the polarization we have to take into consideration  only the Zeeman state with $m=m_s$.

In the next section, we will use relation (\ref{2.9})  to calculate the spatial distribution of atomic polarization  and analyze on this foundation coherent light propagation through ensembles of different densities.

\section{Results and discussion}
\subsection{Atomic polarization}

Expression (\ref{2.9}) allows us to consider atomic clouds with different shapes and with different atomic spatial distributions. We however will further consider mainly model cylindrical clouds with random but (on average) uniform atom distributions along the vector $\mathbf{n}$. For definiteness let us assume that the state $m_s=1$ of the source atom is excited initially. In such a case the source creates a nearly plane right-hand circularly polarized wave in the area of the cloud and the vector of the atomic polarization has only one nonzero component, which we will refer to without index.

Fig. 1 shows the spatial dependence of the absolute value  (Fig. 1a) and phase (Fig. 1b) of the complex quantity $\mathcal{P}^{(+)}(\mathbf{r})$ for different detunings $\Delta=\omega_{s} -\omega _{a}$ of the source probe radiation from exact bare atomic resonance. The calculations were made for a cloud with length  $L=10$ and radius $R=20$. Hereafter in this paper we use the inverse wavenumber of the resonant probe radiation in vacuum $k_0^{-1}=\lambda_a/2\pi=c/2\pi\omega_a$ as a unit of length.  In these units, the mean density of atoms is $n=0.2$. To avoid the influence of boundary effects at the lateral surface of the cylinder as well as diffraction effects caused by the sharp boundary we calculate atomic polarization  $\mathcal{P}^{(+)}(\mathbf{r})$ only for an area near the axis of the cylinder where we  can neglect the dependence of the polarization on $r$. In this area we deal with a quasi one-dimension case.  The polarization depends only on $z$. Our analysis shows that for the considered parameters this take place for the inner portion of the cylinder with  $r\leq 15$. Results shown in Fig. 1 are obtained by averaging of the atomic polarization over the region with radius $r= 10$.

\begin{figure}[th]
\begin{center}
{$\scalebox{1.1}{\includegraphics*{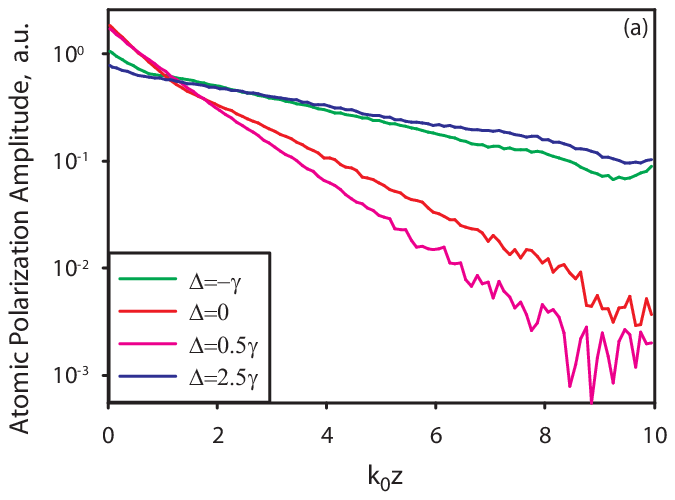}}$ }{$\scalebox{1.1}{%
\includegraphics*{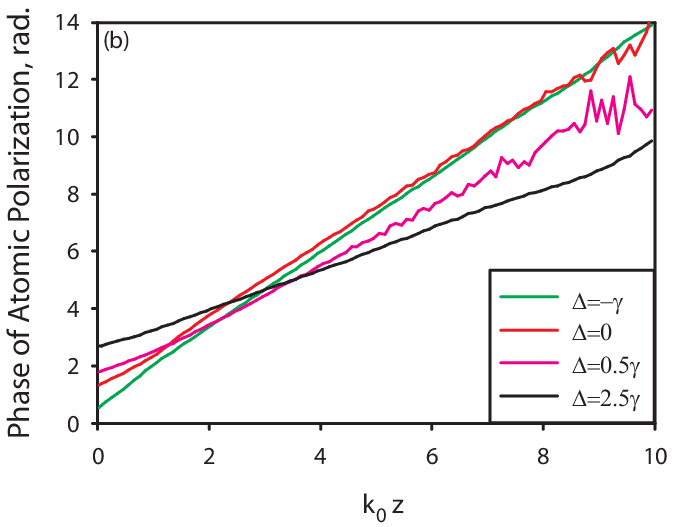}}$ }
\caption{Spatial distribution of atomic polarization. (a) Amplitude of polarization $abs(\mathcal{P}^{(+)}(z))$, semi logarithmic scale; (b) Phase of polarization $arg(\mathcal{P}^{(+)}(z))$. Calculations were made for a cylindrical cloud with length $L=10$ and radius $R=20$, atomic density is $n=0.2$. }
\end{center}
\par
\label{fig1}
\end{figure}
The curves in Fig. 1 were obtained by averaging over the random locations of atoms inside the cloud. The total number of statistical tests was about $6\cdot 10^4$.  Despite such a large number the curves which were not smoothed additionally keep indications of fluctuations. These fluctuations manifest themselves most clearly far from the front edge of the cloud. Here the averaged polarization is extremely small in comparison with the polarization corresponding to any random specific spatial configuration of the ensemble.

In spite of the fluctuations the results shown in Fig. 1 allow us to make several important conclusions about the spatial dependence of the polarization. First, beyond the boundaries and near the ends of the cylinder ($z=0$ and $z=10$) the phase of polarization wave increases linearly. Second, beyond these areas we have a single-exponential decay of the atomic polarization. And last, in the boundary regions with size of about  $1.5\div 2\,\,$ we see peculiarities connected with the fact that atoms located here interact mainly with atoms situated on one side of them, inside the cloud. This causes some modification of the dipole-dipole interatomic interaction. Besides that, the electromagnetic wave reflects from the base edges of the cylindrical clouds. This leads to formation of a standing wave of polarization.  This effect is most evident at the far edge of the cloud ($z=10$) for the wave strongly detuned from resonance (curves corresponded to $\Delta=-\gamma$ and $\Delta=2.5\gamma$). For these waves absorption is small and the amplitude of the reflected wave slightly decreases inside the medium. At the very edge of the cloud we have either a node or an antinode of the standing wave depending on the optical density of the medium. In our case there is a vacuum beyond the cloud. Its refractive index is equal to unity so for large positive detunings we have a node (curve $\Delta=2.5\gamma$ in Fig.1a) at the edge and for negative ones we have an antinode ($\Delta=-\gamma$).

Data shown in Fig. 1 were obtain for several different detunings and one given density of the cloud. Note however that all peculiarities discussed above were observed in our calculation for the full range of considered parameters; that is, for all different detunings and for all considered densities of atoms.

Knowledge of the polarization of the atomic ensemble allows us to make some conclusions about light propagation in it. Three averaged quantities, these being the polarization, the field strength, and the electric displacement are proportional to each other. The coefficients of proportionality for regions away from the boundaries cannot depend on the spatial coordinates because here we deal with a quasi uniform medium. The linear increasing of the phase of the polarization and the single-exponential decay of its amplitude means that in the corresponding area the spatial dependence of the polarization and of the averaged field strength   $\cal E$ are as follows
\begin{eqnarray}
{\cal P }(z)&=&{\cal P}_0 \exp(i(k'+ik'')z); \notag\\
{\cal E }(z)&=&{\cal E}_0 \exp(i(k'+ik'')z).
\label{2.11}
\end{eqnarray}
Here we have taken into account that only one component of each vector is nonzero and that these components depend only on $z$.
The real $k'$ and imaginary $k''$ parts of the wave number can be determined from the decay coefficient and wavelength of the polarization, i.e. from the angles of inclination in the region of the linear dependence of the curves shown in Fig. 1. The results of corresponding calculations are depicted in Fig. 2
\begin{figure}[th]
\begin{center}
{$\scalebox{1.1}{\includegraphics*{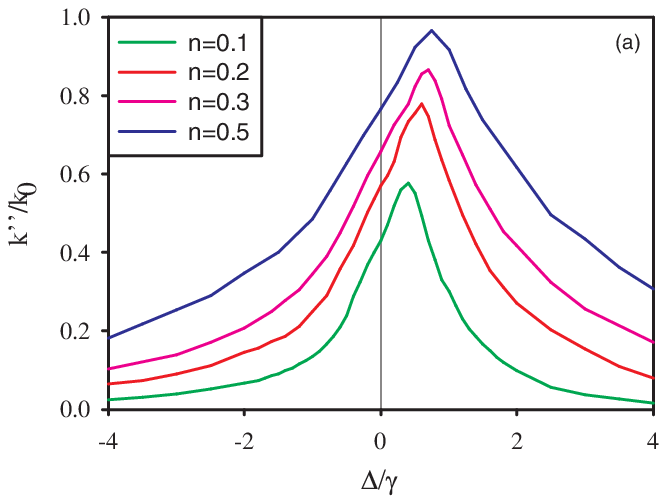}}$ }{$\scalebox{1.1}{%
\includegraphics*{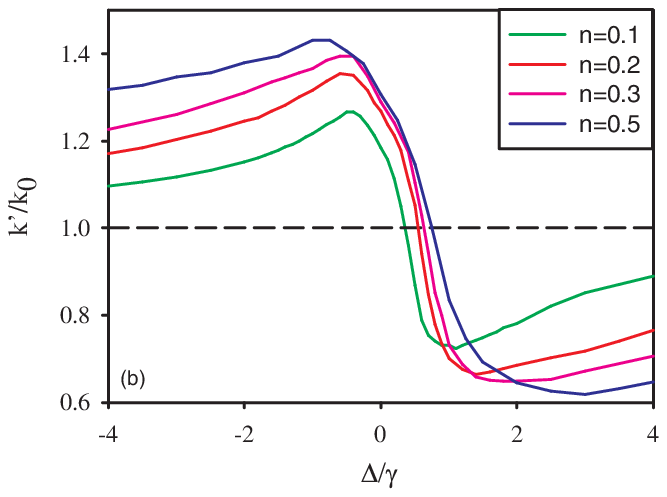}}$ }
\caption{Spectrum of imaginary (a) and real (b) parts of the wave number of a plane electromagnetic wave in atomic ensembles of different density; $k_0=\omega_a/c$ is the wave number of the resonant source radiation in vacuum.}
\end{center}
\par
\label{fig2}
\end{figure}

Fig. 2 shows how interatomic interactions modify the spectral dependencies of absorption and reflection in dense media. In dilute media, both absorbtion and refraction indexes increase linearly with density according to the relation  $k=k_0 + n\sigma(\omega)/2$, where  $n$ is the density, $k$ and $\sigma(\omega)$ are the complex wave number and the complex cross section of light scattering from free atoms.  In the case  when atomic motion and atomic collisions can be neglected the latter gives a Lorentz profile for the absorption coefficient and a corresponding dispersion curve for the refractive index.  The influence of collective effects causes essential distortion of the spectra.  The absorption spectrum is nonsymmetric, there are noticeable shifts of the maximum of absorbtion which, in the considered range of densities, are in the blue wing. The amplitude of the absorbtion increases slowly with density and there is an evident tendency towards saturation. A density increasing from $n=0.2$ to $n=0.5$ leads only to a 25\% increase of maximum absorption, which is much less that under density increasing from  $n=0.1$ to $n=0.2$. Saturation effects in our interpretation connects with level shifts caused by strong dipole-dipole interaction for dense media. These shifts cause also essential nonhomogeneous broadening of the spectral profiles, this being clearly seen in Fig. 2.

Calculation of the real and imaginary parts of the wave number permits us to assess qualitatively the Ioffe-Regel criterium for strong light localization in atomic media. According this criterium, localization can be observed if the transport length of a photon is less than its wave length in the medium. The wave length of the photon is determined by the inverse real part of the wave number  $\lambda/2\pi=1/k'$. The transport length of a photon can be estimated by the absorption coefficient because in the considered media there is no real absorption and attenuation of the coherent component of light connects only with exit of photons from the corresponding mode, i.e. with scattering. Taking into account that the amplitude of the field decreases twice slower the light intensity, the ratio of transport length to wave length can be estimated as $k'/2k''$.  Fig. 3 shows the spectral dependence of this ratio for clouds with different densities.
\begin{figure}[th]
\begin{center}
{$\scalebox{1.1}{\includegraphics*{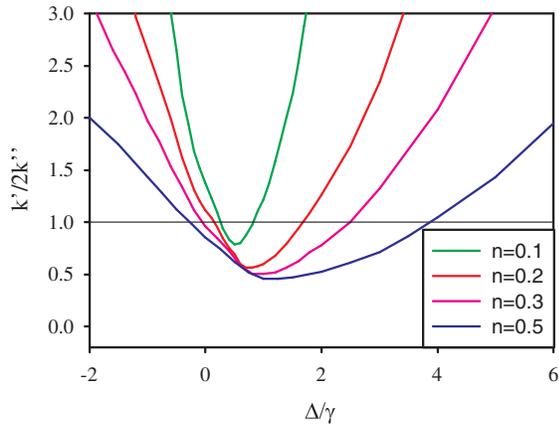}}$ }
\caption{Spectral dependence of the Ioffe-Regel parameter for atomic ensembles with different densities. }
\end{center}
\par
\label{fig3}
\end{figure}

It is seen that even for a density $n=0.1$ there is a region of frequencies where the considered ratio is less than unity. As density increases, the width of the corresponding spectral region also increases. The minimal value of the ratio $k'/2k''$, however, decreases very slowly and shifts into the blue wing. It is also noticeably greater than that predicted for the case of independent scatterers (see for comparison Fig. 3 and Fig. 4 in \cite{Kaiser1}. Here we see directly the influence of the resonant dipole-dipole interaction. Increasing the density results in a decreasing of the portion of atoms which interact effectively with the light at a given frequency.

\subsection{Dielectric constant and atomic susceptibility of dense cold atomic gases}
With knowledge of the complex refractive index we can also calculate the dielectric constant. The latter can be found by the following relations
\begin{eqnarray}
\varepsilon'&=&Re(\varepsilon)=(k'^2-k''^2)/k^2_0 ; \notag\\
\varepsilon''&=&Im(\varepsilon)=2k'k''/k^2_0.
\label{3.2}
\end{eqnarray}
\begin{figure}[th]
\begin{center}
{$\scalebox{1.1}{\includegraphics*{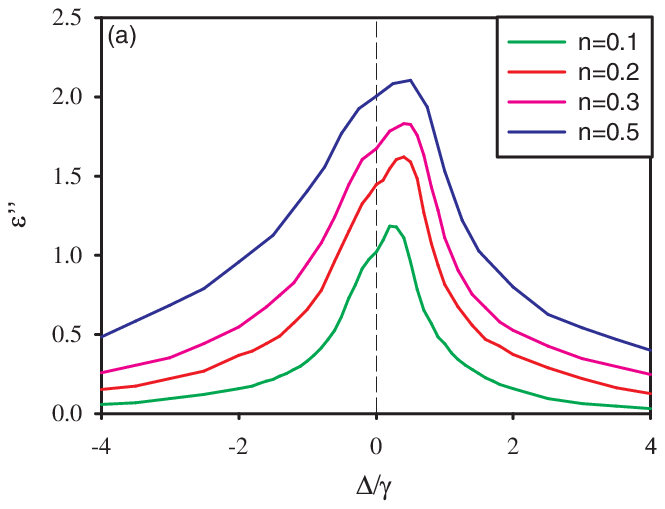}}$ }
{$\scalebox{1.1}{\includegraphics*{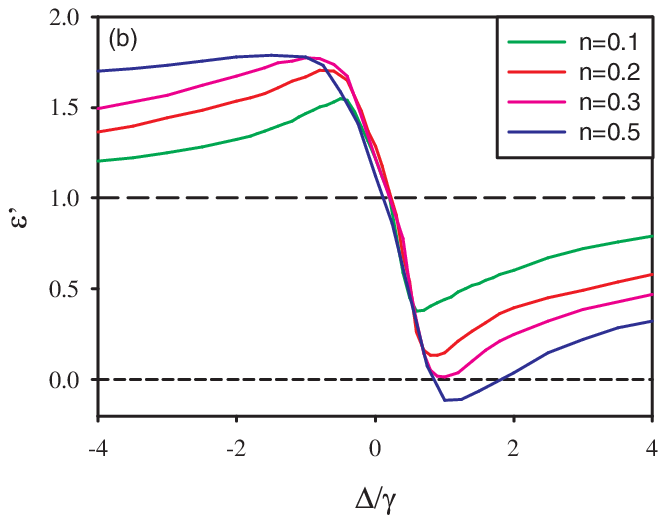}}$ }
\caption{Imaginary (a) and real (b) parts of the dielectric constant for atomic ensembles with different densities. }
\end{center}
\par
\label{fig4}
\end{figure}
In the spectrum of the real and imaginary  parts of the dielectric constant (Fig.4) we see all regularities which were previously observed in the spectrum of the complex refractive index. But there is  one important additional difference. For dense cold atomic media the real part of $\varepsilon$ can be negative at some frequencies. The dipole dynamics is in phase opposition with the driving field. At the considered densities however, and  in the corresponding spectral area the imaginary part $\varepsilon''$ is not negligible and the electromagnetic field keeps wave nature.

The complex refractive index as well as the dielectric constant are used for macroscopic description of the light in media. One of the main characteristics in the microscopic approach is a single atom polarizability  $\alpha$ which is the proportionality factor between averaged dipole moment of an atom and averaged strength of electric field acting on it. The difference between the free atom polarizability and the polarizability in the medium permits us to analyze the mutual influence of atoms in the medium.  The key point in the calculation of the polarizability  is the idea of an effective field acting on the atoms and its distinction from the mean field. In this work we will use the well-known Lorentz-Lorenz formula connecting the mean atomic polarizability and the dielectric constant \cite{BW}.
\begin{eqnarray}
\alpha&=&\frac{3}{4\pi n}\frac{\varepsilon-1}{\varepsilon+2}.
\label{3.3}
\end{eqnarray}
Substituting the known dielectric constant in this equation give us the real and imaginary parts of the polarizability. The results of corresponding calculation are shown in Fig. 5. The main result here is the essential decreasing of the polarizability as the density increases. At higher densities the collective atomic states are distributed over a wider region of frequency and polarizability at a given frequency per one atom is smaller.
\begin{figure}[th]
\begin{center}
{$\scalebox{1.1}{\includegraphics*{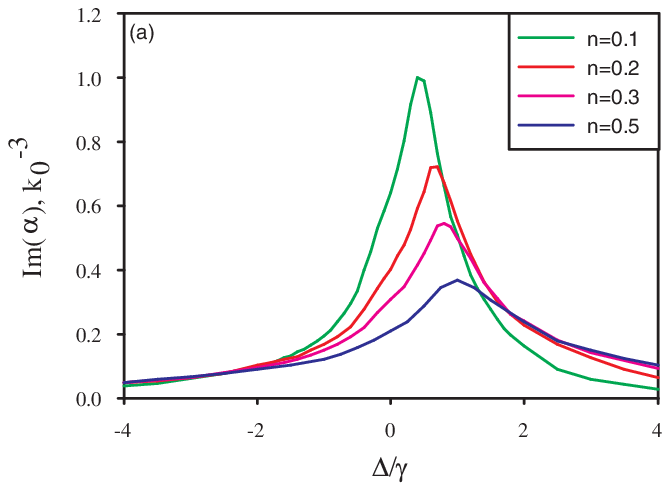}}$ }
{$\scalebox{1.1}{\includegraphics*{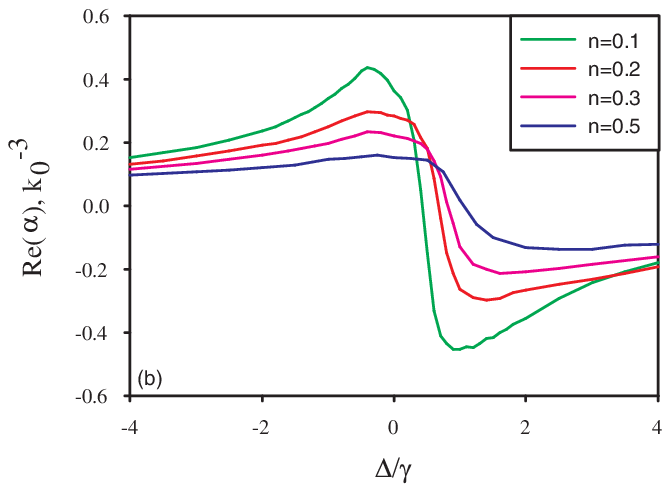}}$ }
\caption{Imaginary (a) and real (b) parts of the atomic polarizability in ensembles with different densities.}
\end{center}
\par
\label{fig5}
\end{figure}

To conclude this section, let consider the quality of the derived dielectric constant and its application in a macroscopic approach. We compare two different results for the total cross section of light scattering from a homogeneous sphere. The first result is obtained in \cite{SKH11} by means of a microscopic calculation.  The second one is the cross section calculated in the framework of the well known Debye-Mie model with the permittivity from this paper.  The result of comparison is shown in the Fig. 6.

\begin{figure}[th]
\begin{center}
{$\scalebox{1.1}{\includegraphics*{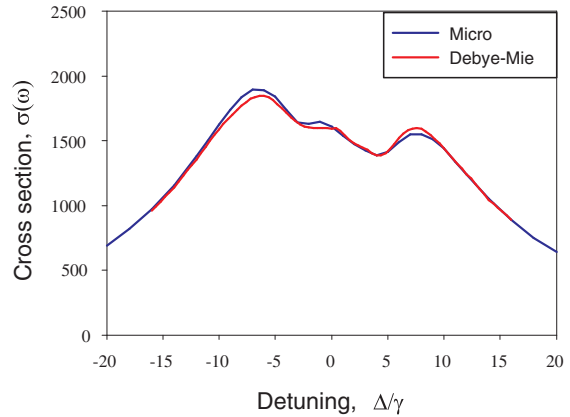}}$ }
\caption{Spectrum of the total cross section of light scattering from a spherical cloud with radius $R=15$. The atomic density is $n=0.2$. The first curve  is the results of consistent microscopic \cite{SKH11} calculations. The second curve is calculated on the basis of Debye-Mie model with the permittivity obtained in this paper. }
\end{center}
\par
\label{fig6}
\end{figure}

The quantitative difference between these two results does not exceed  a few percents. It is very good agreement especially taking into account the approximate numerical determination of the permittivity. We have also noticed that Debye-Mie model is exactly valid for a homogeneous sphere whereas our atomic cloud has boundary regions with  different local permittivity and hence is not completely uniform.

\section{Conclusion}
In the present paper we consider the influence of the resonant dipole-dipole interatomic interaction in dense atomic clouds on their optical properties. Dispersion of the permittivity and atomic polarizability  are determined under different conditions. Atomic clouds with densities up to $n=0.5$ are considered. It is observed that for a dense cloud the real part of the dielectric constant can be negative.

The expression for the dielectric constant found here was used for calculation of the resulting spectrum with that of a previous self-consistent approach \cite{SKH11}. This comparison was restricted to the case of ensembles containing several thousand atoms, but good agreement allows us to use the obtained permittivity for macroscopic calculations in cases when the microscopic approach can not be utilized because of technical difficulties, as indicated in the introduction to this paper.

In this work we also determine the spectral regions, for each considered atomic density, for which the mean free path is smaller than the wavelength, i.e.  we specify conditions when the Ioffe-Regel criterium for strong localization of light in cold dense atomic gases is satisfied. However more definitive conclusions about the possibilities of strong localization need additional study of its direct manifestation in, for instance, the distribution of fluctuations of the transmitted light intensity or in the afterglow delay.

Finally, all calculations in this work were made under the assumption of motionless atoms but for non degenerate gases. In our opinion the developed approach is applicable to the case of quasi-resonant compressible dipole \cite{Kaiser1} or quasistatic electric dipole traps \cite{BWHSK1}. The typical temperature of 30-100 mK achieved for the dense Rb cloud in such traps (see for example \cite{BWHSK1}) is large enough to ignore all effects of degeneracy which can strongly affect light scattering from quantum gases \cite{Gorlitz}. On the other hand, the atomic velocity is sufficiently small here to neglect the Doppler shift (it is several times smaller than the natural width of the excited atomic levels) and to allow us to consider the dipole-dipole interaction as resonant. Averaging over all possible random position of the motionless atoms in our model allows us to take into account the residual motion of real atoms in the traps.

It seems important to further generalize the developed approach to the cases when atomic motion plays a more significant role, for example to the case of hot gases. Such generalization is important for a wider range problems of precision spectroscopy, particularly spectroscopy of selective reflection from the boundary of a dielectric-dense atomic gas \cite{33,Akul,Guo,Bloch,34,Fof1,Fof2}. In this case, however, the dipole-dipole interaction loses its resonant behavior and collisional broadening should be taken into account, along with essential Doppler effect.

\subsection*{Acknowledgements}
We thank Professor D. V. Kupriyanov for fruitful discussions. We appreciate the financial support of the Russian Foundation for Basic Research Grant No. RFBR-10-02-00103a and the United States National Science Foundation (Grant No. NSF-PHY-0654226).

\end{document}